\documentclass[twocolumn]{aastex63}

\usepackage{amsmath}
\usepackage{ulem}
\usepackage{color}

\newcommand{\uvec}[1]{ \hat{\mathbf #1} }

\newcommand{\figsphere}{
\begin{figure*}[ht]
\includegraphics[width=1.\linewidth]{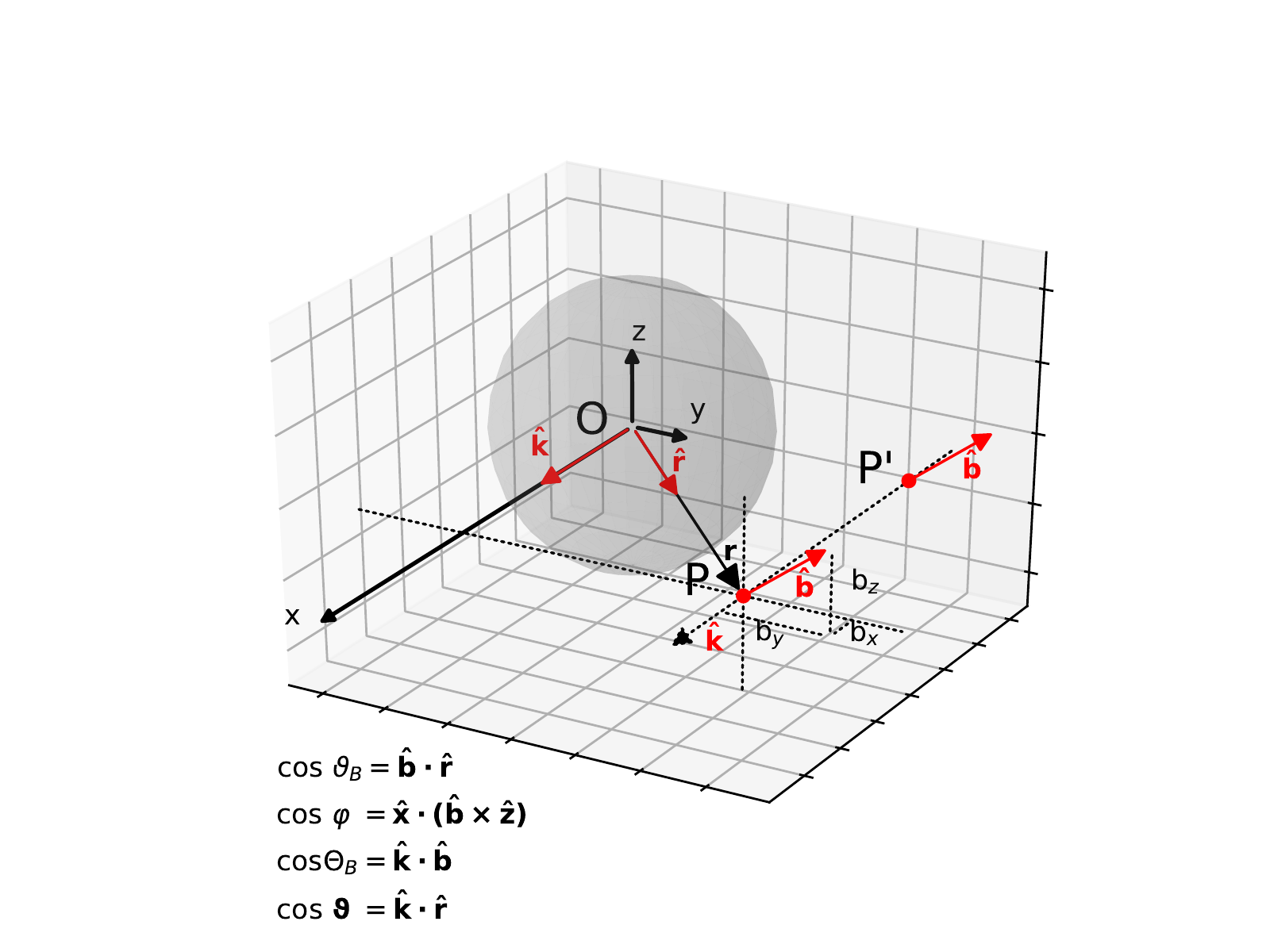}
\caption{The scattering geometry of point $P$ is shown in the observer's frame with 
projections of the magnetic 
field components in this frame.  The line-of-sight 
vector lies along $\mathbf{\hat k}$, and the 
$y$ and $z$ axis are mutually orthogonal, the angle $\Phi_B$ ($=\gamma_B$ in CJ99)  is defined 
by $\arctan \hat b_y / \hat b_z$ if we define the reference direction for linear polarization along the +ve $z-$ axis. The unit vectors 
of interest are marked with red arrows.
}
\label{fig:sphere}
\end{figure*}
}

\shorttitle{Coronal M1 lines}
\shortauthors{Judge et al.}

\begin{document}

\title{On single-point inversions of magnetic dipole lines in the corona}

\correspondingauthor{Philip Judge}
\email{judge@ucar.edu}
\author{Philip Judge, Roberto Casini}
\affiliation{High Altitude Observatory,
National Center for Atmospheric Research,
Boulder CO 80307-3000,
 USA } 
\author{Alin Razvan Paraschiv}
\affiliation{National Solar Observatory, 3665 Discovery Dr, Boulder, CO 80303,  USA}

\begin{abstract}
Prompted by a recent paper by Dima and Schad, 
we re-consider the problem of inferring magnetic properties of the corona using polarimetric  observations of magnetic dipole (M1) lines.  Dima and Schad point to a potential source of degeneracy in a formalism developed by Plowman, which under some circumstances can lead to the solution being under-determined.  Here we clarify the nature of the problem. Its resolution lies in
solving for the scattering geometry using the  elongation of the observed region of the corona. We discuss some conceptual problems that arise when 
casting the problem for inversion in the observer's reference frame, and satisfactorily resolve difficulties identified by Plowman, Dima and Schad. 
\end{abstract}

\keywords{Solar corona; Solar magnetic fields}

\section{Introduction}
\label{sec:problem}

\citet{Plowman_2014} developed  
a method to extract magnetic information from polarized magnetic dipole (M1) emission lines formed in the corona.  In his ``single point inversion'' approach, 
the emergent polarized line profiles 
(measured through Stokes parameters $I,Q,U$ and $V$) are assumed to be dominated by emission from a single  
region along the line-of-sight (LOS)
of thickness $\ell \ll R_\odot$, 
where $R_\odot$ is the solar radius of  
$7\cdot10^{10}$ cm. 

The essence of Plowman's method is to 
use measured $I,Q,U$ and $V$ profiles for
two M1 lines to determine components of the magnetic field within the volume 
defined by $\ell\cdot A$ where $A$
is the projected area of one spatial pixel of the instrument used to measure 
the profiles. From seven independent 
measurements of $I,Q,U$ and $V$ from 
two M1 lines ($Q$ and $U$ containing redundant information, see below), Plowman derived algebraic 
expressions in which seven magnetic and thermal parameters are given in terms of
seven independent observables.   

In a recent assessment of Plowman's work,  
\citet{Dima+2020} identified 
a degeneracy
which can cause the algebraic  solutions
to fail. They 
 argued that, for many pairs of commonly used M1 lines, the 
algebraic solutions are formally 
\textit{undefined}.  Our purpose is to  re-examine  
this problem.   

\figsphere

\subsection{Overview of M1 line formation}

In the quest to measure 
magnetic fields within the corona,
first we must understand the 
origin of the emergent radiation from emitting plasma.  
M1 lines form in the ``strong field limit'' of the Hanle effect \citep[henceforth CJ99]{Casini+Judge1999}.  These lines are generally weak relative to sources of noise \citep[e.g.][]{Penn+others2004}. Thus it is advantageous to 
integrate over the frequency-dependent line profiles, using suitable weights (the $V$ profiles are anti\--symmetric around line center). Here we assume that
all emission comes from a single
homogeneous volume, illuminated by a spectrally flat radiation from below, along a given line of sight with length $\ell$. Then the 
emergent Stokes parameters are (equations 35a-35c of CJ99):
\begin{subequations}
\begin{eqnarray}
	\label{eq:I}
I
&=&C_{J J_0}\ell\,
	\left[1+\textstyle{\frac{1}{2\sqrt{2}}}(3\cos^2\!\Theta_B-1)\,
	D_{J J_0}\,\sigma_J\right], \\
	\label{eq:Q}
Q
	&=&C_{J J_0}\ell\,\,
	\textstyle{\frac{3}{2\sqrt{2}}}\sin^2\Theta_B\cos2\Phi_B\,D_{J J_0}\,
	\sigma_J\;, \\
	\label{eq:U}
U
	&=&-C_{J J_0}\ell\,
	\textstyle{\frac{3}{2\sqrt{2}}}\sin^2\Theta_B\sin2\Phi_B\,D_{J J_0}\,
	\sigma_J\;, \\
	\label{eq:V}
V
&=&-C_{J J_0}\ell\,\frac{\omega_B}{\Delta\omega_D}\cos\Theta_B
	\left(\bar{g}_{JJ_0}+E_{J J_0}\,
	\sigma_J\right)\;.
\end{eqnarray}
\end{subequations}
Here, the Doppler width 
$\Delta\omega_D$ of the line arises from random thermal and 
other motions within $\ell\cdot A$. We  assume that $\Delta\omega_D\gg \omega_B$, 
where $\omega_B$ is the Larmor frequency (CJ99). 
These expressions for Stokes parameters are for
a M1 transition
decaying from atomic 
level $J$ to level $J_0$ (here we 
use $J$ and $J_0$ to identify the unique upper and lower level of the line). 
The resulting frequency-integrated  $I,Q,U$, and $V$ parameters depend 
on quantities ($D_{JJ_0}$, $E_{JJ_0}$, 
$\overline g_{JJ_0}$) determined by the quantum mechanics of the isolated ion, which we assume to be known. There remains six unknown parameters which contain the desired information on the plasma
and magnetic field within $\ell\cdot A$. 

\subsection{Diagnosing the magnetized plasmas}

In attempting to 
derive physical parameters 
from equations~(\ref{eq:I})--(\ref{eq:V}) from 
a set of $I,Q,U,V$ measurements, 
we face several challenges. In these equations the 
unknowns that can be solved for are 
$C_{JJ_0}$, $\ell$, $\sigma_J$, $\sigma_{J_0}$,
$\omega_B/\Delta\omega_D$, $\Theta_B$ and 
$\Phi_B$.
Two parameters can be derived directly from the observed
profiles.  Firstly,
spectrally-resolved line profiles yield the Doppler width 
$\Delta\omega_D$.   Secondly, 
the frequency-integrated ratio
$U/Q$ immediately gives
 \begin{equation}
 \label{eq:U2Q}
  \Phi_B = -\frac{1}{2} \arctan \frac{U}{Q}
  \end{equation}
Two of the five remaining 
parameters ($C_{J J_0}$ and $\sigma_J$) depend on sums and differences of the populations of magnetic sub-states of level $J$. The ``atomic alignment" $\sigma_J$
has an implicit dependence on the magnetic field 
geometry (CJ99, section 3.3) 
 generated by the angle between the unit vectors $\hat{\mathbf r}$ from Sun center to the radiating point and $\hat{\mathbf b}$
 (see Figure~\ref{fig:sphere}).  This angle $\vartheta_B$
 (with cosine 
 $\mathbf{\hat{ b }
 \cdot \hat{ r }}$)
 appears only implicitly
 in equations (\ref{eq:I}) to 
(\ref{eq:V})
 through the term $\sigma_J$. Clearly, $\vartheta_B$ varies 
 with position $x$ along the LOS. 
 
The  three magnetic parameters that might be derived from observed  Stokes parameters are related to the vector magnetic field, two angles 
$\Theta_B$ and
$\Phi_B$ which define $\hat{\mathbf{b}}$, and 
$\omega_B$, where $$\hbar\omega_B=\mu_B B,$$ with $\mu_B$ the Bohr magneton, and $B$ is the magnetic field strength.   
Angles $\Phi_B$ and $\Theta_B$  (equations 39a-39d of CJ99) are defined in the observer's frame
(see Figure~\ref{fig:sphere}).  $\Phi_B$ is
the azimuthal angle of the magnetic field vector projected on to the plane-of-sky (POS), and $\Theta_B$ is the angle between the line-of-sight vector $\hat{\mathbf{k}}$ and the magnetic field vector $\hat{\mathbf{b}}$.

The common factor $C_{JJ_0}$
is the coefficient for isotropic emission for the line intensity:
\begin{eqnarray}
C_{J J_0}
&=&\frac{\hbar\omega}{4\pi}\,N_J A_{JJ_0}\;,
\end{eqnarray}
where $N_J$ is the population density of the upper level $J$
(i.e. the sum over all magnetic sub-states), and $A_{JJ_0}$ the Einstein A-coefficient.   Both $N_J$ and $\sigma_J$ depend on the densities and temperatures of the plasma through collisional terms in the statistical equilibrium equations. 
$D_{JJ_0}$, and $E_{JJ_0}$ 
\citep{Casini+Judge1999}
depend only on $J,J_0$, and the effective Land\'e factor of the transition $J\rightarrow J_0$ is
\citep{Landi}:
\begin{equation}
{\bar{g}_{JJ_0}}=
    \frac{1}{2}( g_J+g_{J_0}) +
    \frac{1}{4}( g_J-g_{J_0})
\left[J(J+1) - J_0(J_0+1)\right],
\label{eq:gbar}
\end{equation} 
Here, $g_J$ and $g_{J_0}$ are the Land\'e g-factors for the splitting of levels $J$ and $J_0$, properties of the isolated ions assumed known from experiment and/or theory. 

\subsection{Plowman's algebraic inversion}

Plowman uses 
observations of $I,Q,U$, and $V$ from two M1 lines
from the same ion
of Fe$^{12+}$. Along with ions of the carbon iso-electronic sequence,  silicon-like Fe$^{12+}$ has a 
$^3P_{2,1,0}$ set of ground levels between which 
the $J=1 \rightarrow J_0=0$ and 
$J=2 \rightarrow J_0 = 1$ M1 transitions 
occur. Anisotropic irradiation 
of these coronal ions by
radiation from the solar 
surface causes 
$\sigma_J$ to be non-zero,
hence leading to linear polarization (equations 
\ref{eq:Q} and \ref{eq:U}). In using pairs of lines from the same ion, one can eliminate
the need to include abundances and 
ionization fractions when computing $N_J$. In short, one has eight observed (frequency-integrated) 
Stokes measurements from which eight unknowns might be
derived, which in principal 
admits algebraic solutions.

Three of these unknowns specifying the magnetic field vector ($\omega_B,\Theta_B,\Phi_B)$ are common to the two lines. The Doppler width  $\Delta\omega_D$ is readily derived from the observed 
line width, leaving the alignments $\sigma_J$, $\sigma_{J_0}$ and column density $N_J \ell$ and
$N_{J_0} \ell$ as  remaining line-dependent unknowns. We therefore have a total of seven unknowns.
Further, the ratio of equations (\ref{eq:U}) and 
 (\ref{eq:Q}) 
 for each line yields just one unknown 
  from each measurement of $Q$ and $U$, independent of $I$ and $V$.   
  The summed squares of  equations (\ref{eq:U}) and 
 (\ref{eq:Q}) for each line 
 yields the magnitude $L_i$ of the
  linear polarization
   of line $i$ 
   through 
$L_i^2 = Q_i^2 + U_i^2$. 
As equation~(\ref{eq:U2Q}) applies to both lines, 
only seven of the observed Stokes parameters can be treated as independent measurements, these are
\citep[][section 2]{Plowman_2014}:
$$I_1,L_1,V_1,I_2,L_2,V_2, (U_1/Q_1 = U_2/Q_2).$$ 
The essence of the method is to thus to determine the seven parameters 
$$
\omega_B,\Phi_B
\pm n{\pi/2},\Theta_B,
N_{J_1}\ell, \sigma_{J_1},
N_{J_2}\ell, \sigma_{J_2},
$$
algebraically from the remaining observables.  If the method is applied to more than  2 lines, the problem 
becomes one of minimizing a goodness-of-fit instead of an algebraic solution. 

\section{Degeneracies}

 The problem identified by \citet{Dima+2020} arises 
when a derived atomic quantity $F_{JJ_0}$ is zero for both of the observed lines:
\begin{eqnarray}
    \label{eq:Fdima}
    F_{JJ_0}&=& \sqrt{2}\frac{E_{JJ_0}}{D_{JJ_0}} - \overline g_{JJ_0}\\
    &=& \frac{3}{4} \left[ 
   J(J+1) - J_0(J_0+1) - 2
    \right] (g_J-g_{J_0})
\label{eq:Fexpanded}
\end{eqnarray}
The latter equality applies to $\Delta J =0,\pm1$ \citep{Dima+2020}.  In passing we note that although most M1 lines 
are between levels of the same (ground) term where
$\Delta J= \pm1$, 
there exist other M1 coronal transitions 
with $\Delta J=0$. 
For example C-, Si- and O- and S-like ions 
possess M1 (and E2) lines
with $J=J_0$, for example 
between the $np^2~^1D_2$ and $ 
np^2~^3P_2$ levels, $n=2,3$.
A particular example is the 
$3p^2~^1D_2 \to  
3p^2~^3P_2$ transition of Si-like \ion{Fe}{13} at 338.85 nm, observed during the 1965 total eclipse by 
\citet{Jefferies+Orrall+Zirker1971}.  \citet{Jordan1971} reports the same transition in S-like 
\ion{Ni}{13} at 212.6 nm, obtained during the 1970 total
eclipse from a rocket spectrometer. Such transitions occur only when 
LS coupling breaks down, the  level wavefunctions become instead mostly mixes 
of the two LS-coupled levels involved.  

Clearly $F=0$ when $J=1, J_0=0$ or when $g_J = g_{J_0}$. Table~1 of \citet{Dima+2020}
lists important M1 lines from the C- and Si-like isoelectronic sequences
for which $F_{10}=0$ and, in LS-coupling, 
$g_{2} = g_{1}$.

Algebraic elimination of all unknowns except $\Theta_B$
from 
equations (\ref{eq:I}) to 
 (\ref{eq:V}) yields a solution for $\sin^2 \Theta_B$ in terms of the Stokes measurements and atomic parameters including $F_i$.  
 To find the algebraic solutions we assume 
that products $D_i\sigma_i$ for both lines have the same sign  (this must be the case physically unless the linear polarization is modified in a multi-level atom via other radiative or collisional transitions).  In practice this assumption corresponds to the situation where the atomic alignment is determined by optical pumping of similarly
anisotropic radiation 
through the polarizability factor $D_i$
for both lines. 
Then we find\footnote{See the appendix.  The wavelength dependence arises because the Doppler width of the spectral line $\Delta\omega_D$ 
 is proportional to
 wavelength.  Equation (11) of  \citet{Dima+2020} contains this dependence only implicitly. However their definition of $V$ (their equation 8) is not in the same 
 units as $I,Q,U,$ and must be used in
 their equation (11).} 
 \newcommand{\st}{\sin^2\Theta_B}
 \begin{eqnarray}  \nonumber 
\left [  {\bar{g}_{1}\lambda_{1}} 
        (I_1 \pm L_1)V_2 -
         {\bar{g}_{2}\lambda_{2}} 
    (I_2 \pm L_2)V_1
\right ]\st
 \\ = \pm
 \frac{2}{3} (
\lambda_1 F_1 L_1 V_2 -  \lambda_2 F_2 L_2 V_1) 
\label{eq:soln}
\end{eqnarray}
This equation differs from that of \citet{Dima+2020} by including explicitly the wavelengths $\lambda_i$ instead of incorporating them into a revised  definition of $V$.
If the term in square brackets is non-zero, two cases can be  examined in terms of the atomic quantities $F_i$, independent of consideration of the measurements:
\begin{itemize}
    \item $F_1=F_2=0$: The RHS is identically zero, thus the LHS of this equation is zero. Either the term in brackets [] is zero and/or 
    $\sin^2 \Theta_B=0$.   
    If the bracketed term [] is zero, the [] term gives an equation linking all measurements $I,L,V$ of both lines, then there are fewer observables than model parameters. When [] is non-zero, 
this implies $\sin^2 \Theta_B=0$.  But when this is true,  there can be no linear polarization and 
    both $L_i$ must be  zero
    no matter the measured values.
    Then as emphasized by \citet{Dima+2020}, no  solution other than  $\sin^2\Theta_B=0$ is possible.
    
    \item Either of $F_1$ or $F_2$ or both are non-zero. The RHS of the equation is non-zero, so that both the bracket [] and $\sin^2\Theta_B$ are non-zero. For a given set of measurements, two solutions are possible through the $\pm L_i$ terms, including at least one that is physically acceptable, compatible with the reality condition $0 \le \sin^2\Theta_B \le 1$. 
\end{itemize}
The existence of  solutions for $\sin^2\Theta_B$
can also be related to values of  \textit{observed} parameters $I_i, L_i, V_i$, for given, known non-zero values of at least one of the $F_i$.  
The $I_i, L_i, V_i$ measurements have intrinsic uncertainties, and so we must consider the statement 
``measurement $S_i=0$'' to mean that the observed value, within measurement
uncertainties, is compatible with 
zero. 
\begin{itemize}
    \item If all $V_i=0$ but $L_i \ne 0$, 
then, even though $\Phi_B$ can be defined (equation~\ref{eq:U2Q}), 
$\sin^2\Theta_B$ is undetermined. 
\item If all $L_i=0$ but 
at least one of the $V_i$ are
non-zero, the only solution
possible is $\sin^2\Theta_B=0$, with $\Phi_B$ undetermined.
\end{itemize}
Both of these algebraic cases 
reflect the intuition that 
non-zero  circular and linear polarization values are required to infer magnetic field properties.

We conclude that the formulation of this problem, originally based on
the notion that from $N$ measurements that are independent one can derive $N$ parameters, leads to difficulties arising when 
na\"ively applying observed 
values in say equation~(\ref{eq:soln}). Internal dependencies in the model 
equations (\ref{eq:I}) to 
 (\ref{eq:V}) show that the 
 observations cannot be all independent, when there are hidden symmetries. One of these symmetries found by \citet{Dima+2020} occurs when $F=0$.


\section{Commentary}

\subsection{Incomplete formulation of the problem}
\label{subsec:incomplete}

When one or more of the values $F_i \ne 0$, Plowman's (2014) 
problem recovers three magnetic variables $\Theta_B$, $\Phi_B$, $B$, and the populations $N_J$
and alignments $\sigma_J$ for each transition's upper level $J$, from the Stokes parameters of two M1 lines. In recovering 
the signs of $\sigma_J$, 
a well-known ambiguity 
in $\Phi_B$ 
of $n\frac{\pi}{2}$ with $n$ any integer, 
is reduced to 
$n{\pi}$ (see equations (\ref{eq:Q}) and (\ref{eq:U})) because two of the angular  quadrants for $\Phi_B$ are eliminated \citep{Judge2007}.  
 Subject to these ambiguities, the  magnetic field vector can be regarded to be \textit{known}. In principle, a map of $\mathbf{B}$ in the plane-of-sky ($x=0$, say) might 
then be constructed.  Such maps
were made commonly in early 
coronagraphic studies of 
just $I,Q,U$ coronal data 
\citep{Querfeld1977,Querfeld,Arnaud}.

However, we have 
become concerned that these algebraic solutions, written explicitly in terms of angular variables 
$\Theta_B$ and $\Phi_B$  defined in the 
observer's frame,
appear to be \textit{independent of the scattering geometry}.  
This point can be appreciated by inspection of Figure~\ref{fig:sphere}.
Any  unit magnetic field vector fixes 
the values of $\Phi_B$ and $\Theta_B$ for all values of $x$ along the LOS.  The only information on the $x$ 
coordinate in the algebraic solutions 
is therefore encoded only 
in the atomic alignments
$\sigma_J$ and $\sigma_{J_0}$. 
The level populations $N_J, N_{J_0}$  determine just the 
total number of emitted  photons.

Suppose that we know by independent means that the bulk of the emission comes from values of $x$ where $|x| \ll R_\odot$, and we 
have in hand the 
solutions for the seven variables from Plowman's method.  Two questions then 
arise: are the alignment factors derived physically acceptable? Are they compatible with $x\approx 0$ to within 
uncertainties? 

This thought experiment suggests that 
the data might  
be used also to constrain the coordinate $x$ of the emitting plasma.  With $\Phi_B$ 
determined modulo  $n \pi$ from equation~(\ref{eq:U2Q}), and $\Theta_B$ 
from a successful application of 
equation (\ref{eq:soln}),
one can imagine emission originating from different points along the $x$ axis, 
for fixed $y$ and $z$. The 
magnitudes and signs of
alignments are determined in part by the value of $\cos \vartheta_B =
\hat{\mathbf{b}}\cdot \hat{\mathbf r} $.
Now  
$\hat{\mathbf r}$ varies
according to the geometry independent of
the fixed value of $\hat{\mathbf{b}}$,
therefore we see that the atomic alignments implicitly contain information about the LOS coordinate
$x$ of the plasma. 

As originally conceived by one of us (PGJ), the method developed by Plowman tacitly  assumed that the plasma emission would arise from 
regions where 
$|x| \ll R_\odot$.   This  assumption,
also adopted in the earlier 
work \citep{Querfeld1977,Querfeld,Arnaud}, 
can only be  weakly 
justified, noting that  the plasma pressure scale height $h$  $\sim 0.06R_\odot$ in the inner corona.  Regions where  $|x| \ge \sqrt{R_\odot h} \approx 0.25 R_\odot$ will typically be too tenuous to contribute significantly to 
 line emission. With 
this assumption, the geometry is fixed  with 
$\cos \vartheta=\hat{\mathbf{k}} \cdot \hat{\mathbf{r}} \approx 0$.  The  emission lines 
we observe
are scattered by  $\vartheta\approx 90^\circ$ 
towards the observer.   If we choose to accept these conditions,
equations (42) and (44) of \citet{Casini+Judge1999} read 
$$ \cos \Theta_B = \sin\theta_B \cos\varphi_B$$
and 
$$
\sin\Theta_B\cos\Phi_B 
= \cos\vartheta_B$$
which, given 
a particular solution to Plowman's problem, 
are sufficient to solve for $\vartheta_B$ and $\varphi_B$ to define the  geometry in the solar rest frame. 

\subsection{An explicit  formulation}

In  hindsight, the
inversion scheme
of \citet{Plowman_2014} is seen as 
an incomplete 
determination 
of plasma and magnetic properties from Stokes data.  
In seeking the minimal 
set of seven parameters $N_J, N_{J_0}, \sigma_J, \sigma_{J_0}, \Theta_B,\Phi_B$,
and $B$ from seven independent
measurements $S_i$, the 
scattering geometry is not explicitly 
treated.  Yet we argued in
the previous subsection that 
such information is implicitly 
contained in the atomic alignments.  

These difficulties  have prompted us to reformulate 
this ``inversion problem'',
to 
build a database of 
Stokes parameters computed from single points along the line-of-sight  
\citep{Judge+Paraschiv2021}. Stokes parameters computed within the grid are sought to 
match observations through
a goodness-of-fit metric.
The $y$-coordinate (i.e., astronomical 
elongation) is used as an 
observable, allowing us solve for $x$ values  matching observation and theory.   Only two dimensional searches ($x$ and $y$) are needed
in the database to find the optimal solutions, 
because the statistical equilibrium equations for the radiating ions are invariant to rotation by an angle $\alpha$ around the  $x$-axis
for spherically symmetric 
radiation from the solar surface.  Thus, the 
Stokes parameters $Q$ and $U$ seen by an observer can simply be rotated through an angle $-2\alpha$ prior to seeking 
solutions in the database's 2D plane.  

How does this explicit approach relate to  that of 
\citet{Plowman_2014}?  Both seek solutions compatible with observations, both have intrinsic ambiguities \citep{Judge2007}.  The  difference is in using the  observed $y$-coordinate 
and a search along $x$ to fix
the scattering geometry.
As we show below, the redundancy problem 
identified 
by \citet{Dima+2020} then vanishes. 
Independent of 
values of $F_{JJ_0}$, in each identified solution, all angles in the solar reference frame are known
(i.e., $\hat{\mathbf{k}}\cdot \hat{\mathbf r}$, 
$\hat{\mathbf{k}}\cdot \hat{\mathbf b}$, and 
$\hat{\mathbf{b}}\cdot \hat{\mathbf r}$; see Figure~\ref{fig:sphere}).  
 The extra numerical work in the database approach is minimal.
\subsection{$F=0$ and the explicit method}

Is the 
``$F=0$ redundancy problem'' of  \citet{Dima+2020} then common to both approaches? Or, does the inclusion of the $y$-coordinate and use of alignment factors to solve for $x$  avoid this problem?  The answers are no and yes respectively. 

\subsubsection*{Proposition} In solving for the geometry using
the additional information in the atomic alignments, the values
of the $F_i$ play no role in 
the determination of $\Theta_B$.

\subsubsection*{Proof}

Consider the geometry
of Fig.~\ref{fig:sphere}, and make no assumptions about
$F_i$.  For simplicity  
assume that the unit vector $\hat{\mathbf{b}}$ of the magnetic field
is fixed along any given $y_0$.  
First we demonstrate that 
the alignment factors are 
simple functions of the $x$-coordinate of the emitting volume $\ell \cdot A$.  

First consider the dependence on $x$
of the angles $\vartheta,\vartheta_B, \phi_B$ in the solar reference frame.  
$\vartheta= \arctan y_0/x$ is clearly a single valued function
of $x$ for any $y_0$. 

The alignment factors are indeed 
simple functions,
when generated by unpolarized, and anisotropic but cylindrically symmetric photospheric radiation. In this case they are proportional to the anisotropy factor\footnote{This equation
neglects limb darkening, but this
is not essential to the present argument.} 
\begin{equation}
\label{eq:J20}
    \frac{J_0^2}{J_0^0} 
    = \frac{1}{4\sqrt{2}}
    (3\cos^2\vartheta_B-1)
    (1 + \cos\vartheta_M)\cos\vartheta_M,
\end{equation}
where $\vartheta_M$ is the half-angle
defining the cone of solar irradiation (this is equation~(31) of \citealp{Casini}). 
The dependence of $\sigma_J$ on
$\vartheta_B$ leads to the well-known Van Vleck effect (e.g., \citealt{Sahal}).  From Figure~\ref{fig:sphere} we see
\begin{equation}\label{eq:cvtb}
\cos\vartheta_B = \uvec{b}\cdot\uvec{r} = b_x\, x + b_y y_0
\end{equation}
which for fixed $\uvec{b}$ is a single-valued function of $x$ for $0\le \vartheta_B \le \pi$. Finally,
the anisotropy factor (equation~\ref{eq:J20}) for a fixed 
$y_0$ is  a  function
of $x$, because (CJ99 equation 
29 using $R_\odot=1$)
\begin{equation}\label{eq:svtm}
    \sin \vartheta_M
    = \left( 1 + h\right)^{-1},
\end{equation}
where in our notation 
$h = |\sqrt{ x^2 + y_0^2}|-1$, 
evidently is a function of $|x|$.
So we can write, symbolically, the angle dependencies 
for any measured elongation $y_0$, as follows:
$$
\vartheta\{x;y_0\},\  \vartheta_B\{x;y_0\},\  \vartheta_M\{|x|;y_0\} 
$$
where $\{\}$ implies a unique functional dependence, for each observed $y_0$. 
Equations (\ref{eq:J20}),
(\ref{eq:cvtb}),
and 
(\ref{eq:svtm}) show that 
$\sigma_J$ and $\sigma_{J_0}$
are single functions of $x$
(not just $|x|$).

To complete the proof we must relate these 
dependencies of angles in the solar frame to the angles
$\Phi_B$ and, in particular, 
$\Theta_B$ in the observer's frame.  
$\Phi_B$ is determined modulo $\pi/2$ radians  directly from the observed $Q$ and $U$ through equation~(\ref{eq:U2Q}). It is further determined to modulo 
$\pi$ when the signs of 
the alignments are known. 

The geometric quantities derived by the 
explicit method are  
\begin{displaymath}
\Phi_B,\
\vartheta\{x;y_0\},\ 
\vartheta_B\{x;y_0\},\
\sigma_J\{x;y_0\},\
\sigma_{J_0}\{x;y_0\}, 
\end{displaymath}
and these functional dependencies are multi-valued (e.g. modulo $\pi$ for $\Phi_B$) but otherwise  non-degenerate.
Now, applying the spherical 
trigonometry equations (42) and (44) of \citet{Casini+Judge1999}, we see that, given the above angles, we can eliminate $\phi_B$ and solve for $\Theta_B$, which therefore is also 
functionally dependent on $x$.  The method, in solving for $x$, also solves for 
$\Theta_B$ without use of 
equation~(\ref{eq:soln}). 
No information on $F$ was required in this argument.  This completes the proof.

The primary assumption we have made is that the 
alignments are proportional 
to the anisotropy factor (equation~\ref{eq:J20}).  Under most conditions this in the corona, this is a reasonable assumption (see also the discussion by \citealp{Judge2007}). Indeed 
interesting new physical processes could be studied,
such as other sources of anisotropy in the SE equations, if
this were not the case. 

\acknowledgments
The National Center for Atmospheric Research and National Solar Observatory are sponsored by the National Science
Foundation. We thank a referee for
helpful checks of algebra. 

\vphantom{H}
\vskip 24pt
\appendix

\section*{Solution for $\sin^2\Theta_B$}

\renewcommand{\theequation}{A\arabic{equation}}

Define for each spectral line $i$ the quantities
$
\Delta_i = \frac{D_i \sigma_i}{\sqrt{2}}, \ \ \ s=\sin\Theta_B  $.
Then using 
equations (1a)-(1c), and omitting the factor $C_i$ for notational economy, we can write
\newcommand{\sst}{\frac{3s^2}{2}}
\newcommand{\modd}{|\Delta_i|}
\begin{eqnarray}
I_i&=&1 + \Delta_i - \sst\Delta_i \\
L_i&=&\sst|\Delta_i|  \label{eq:li}
\end{eqnarray}
Now define sums and differences of the two 
positive definite observed quantities 
$I_i$ and $L_i$, in terms of
$\modd$, taking into account the two signs taken by 
$\Delta_i$.  
\begin{equation}
      \text{For } \Delta_i > 0 
      \begin{cases}
     I_i+L_i = 1 + \modd \\
     I_i-L_i = 1 + \modd -3s^2     \modd,
    \end{cases}      \label{eq:delpos}
\end{equation}
\begin{equation}
      \text{and } \Delta_i < 0 
      \begin{cases}
      I_i+L_i = 1 - \modd + 3s^2 \modd\\
I_i-L_i = 1 - \modd, 
    \end{cases}       \label{eq:delneg}
\end{equation}
but the two solutions in (\ref{eq:delpos}) and (\ref{eq:delneg}) containing the 
$s^2$ terms are redundant with the others when we recognize that 
$3s^2 |Delta_i|= 2L_i$ (equation~\ref{eq:li}). 
Adding the Stokes $V$ measurements using equation (1d) we have 
\begin{equation}
     V_i \propto \lambda_i B \cos \Theta_B \left \{
    {\overline g_i} (1 + \Delta_i) + F \Delta_i \right \}
\end{equation}
where the wavelength $\lambda_i$ of transition $i$ enters through the Doppler width of the lines ($\Delta \omega_D^{-1}$ in wavelength units and with a change of sign). Then the ratio of  $V$ signals 
needed to yield  equation~(\ref{eq:soln}) 
from two lines $i$ and $j$ becomes 
\begin{equation}
    \frac{V_i}{V_j}
    = \frac{\lambda_i}{\lambda_j} \frac{ \left \{
    {\overline g_i} (1 + \Delta_i) + F \Delta_i \right \}}{\left \{
    {\overline g_j} (1 + \Delta_j) + F \Delta_j \right \}
    }. 
\end{equation}
An equation for $s^2=\sin^2\Theta_B$ can be written in terms only of observables, substituting for
$\Delta_i$ and $1+\Delta_i$ 
using equations~(\ref{eq:li})
(\ref{eq:delpos}) and (\ref{eq:delneg}), taking into account the signs of $\Delta_i$:

\begin{equation}
    \frac{\lambda_jV_i}{\lambda_iV_j}
    = 
    \frac{
    \overline g_i(I_i\pm L_i) \pm \frac{2F_iL_i}{3\sin^2\Theta_B}}
    {\overline g_j(I_j\pm L_j) \pm \frac{2F_jL_j}{3\sin^2\Theta_B}
    }.
\end{equation}
On multiplying by $3\sin^2\Theta_B/2$ and re-arranging we arrive at  equation~(\ref{eq:soln}).

\bibliography{biblio}
\bibliographystyle{aasjournal}

\end{document}